\documentclass[11pt, a4paper, logo, copyright, nonumbering]{deepmind}

\usepackage[authoryear, sort&compress, round]{natbib}
\usepackage{amsmath}
\usepackage{amsfonts}
\usepackage{siunitx}

\usepackage{caption}
\DeclareCaptionLabelFormat{adja-page}{\hrulefill\\#1 #2 (\emph{continued})}

\newcommand{\lpr}{\left(}
\newcommand{\rpr}{\right)}

\newcommand{\textcite}{\cite}

\title{Space is a latent sequence: Structured sequence learning as a unified theory of representation in the hippocampus}

\correspondingauthor{dileepgeorge@deepmind.com}

\author{Rajkumar Vasudeva Raju, J. Swaroop Guntupalli, Guangyao Zhou, Miguel L\'azaro-Gredilla and Dileep George}

\begin{abstract}
Fascinating and puzzling phenomena, such as landmark vector cells, splitter cells, and event-specific representations to name a few, are regularly discovered in the hippocampus. Without a unifying principle that can explain these divergent observations, each experiment seemingly discovers a new anomaly or coding type.  Here, we provide a unifying principle that the mental representation of space is an emergent property of latent higher-order sequence learning. Treating space as a sequence resolves myriad phenomena, and suggests that the place-field mapping methodology where sequential neuron responses are interpreted in spatial and Euclidean terms might itself be a source of anomalies. Our model, called Clone-structured Causal Graph (CSCG), uses a specific higher-order graph scaffolding to learn latent representations by mapping sensory inputs to unique contexts. Learning to compress sequential and episodic experiences using CSCGs result in the emergence of cognitive maps - mental representations of spatial and conceptual relationships in an environment that are suited for planning, introspection, consolidation, and abstraction. We demonstrate that over a dozen different hippocampal phenomena, ranging from those reported in classic experiments to the most recent ones, are succinctly and mechanistically explained by our model. 
\end{abstract}

\begin{document}

\maketitle

\section{Introduction}

The hippocampus is known for its role in episodic memory, map-like spatial representations, relational inference, and fast learning -- a seemingly disparate set of requirements. Simultaneously, hippocampal cells are categorized into a wide variety of types based on their firing patterns ranging from place cells, splitter cells, time cells, lap cells, event specific representations and exhibit a variety of remapping phenomena in response to environmental changes \citep{Kubie2020-bc}. These phenomena often get characterized using Euclidean spatial concepts such as object vector cells \citep{Bicanski2020-lt}, landmark vector cells \citep{deshmukh2013influence}, and distance coding \citep{Sarel2017-vs,deshmukh2013influence}, without a coherent underlying explanation, and remain unresolved with other phenomena like splitter cells \citep{Dudchenko2014-ws,Ainge2007-gq,Ainge2007-fb} and event-specific representations \citep{sun2020hippocampal}. Could these divergent requirements and myriad phenomena be explained using a simple set of principles that are computationally grounded, implemented, and easy to understand? Here we show that treating \emph{space as a sequence} can resolve many of the divergent phenomena ascribed to spatial mapping, and help clarify the connections between spatial, temporal, abstract, and relational representations in the hippocampal complex. 

Treating space as a sequence is a necessity for humans and other animals because they do not have a global positioning system that enables direct sensing of location coordinates. 
Consequently, they need to acquire and abstract the concepts of locations and space from purely sensory-motor experience. However, sensations from the world are aliased and do not convey locations directly -- identical sensations can occur at multiple locations or in different sequential contexts. To develop internal space-like maps from these aliased sensations (as illustrated in the sketch in Fig. \ref{fig:CSCG-concepts}A), the learning agent has to appropriately split or merge  sensations based on sequential contexts \citep{niv2019learning, Plitt2021-gc}. Our model, clone-structured causal graph (CSCG), tackles this problem by learning different latent states (called clones) to represent the same observation in different sequential contexts \citep{dedieu2019learning, George2021-qt, Eichenbaum2004-aq}, merging or splitting them as necessary. In CSCGs, allocentric "spatial" representations naturally arise from higher-order sequence learning on egocentric sensory inputs, without making any Euclidean assumptions, and without having locations as an input. An organism or an agent can utilize a CSCG for navigation, foraging, context-recognition, and shortcut finding without having to explicitly compute place fields, or having to decode locations.

Importantly, our model suggests that place field maps need to be interpreted carefully because they overlay sequential responses on to Euclidean maps. Directly characterizing the place field maps in terms of spatial and Euclidean concepts could be a source of anomalies because the underlying phenomena are inherently sequential and dynamic \citep{Warren2019-bp}. In contrast, CSCG explicates how the learning of sequential contexts gives rise to spatial representations that an agent can use to drive behavior without explicitly representing location coordinates. CSCGs predict the conditions under which place fields are expected to change in response to visible or invisible environmental changes, and when they do not, resolving a variety of phenomena with a simple principle. 

\section{Model}
We consider experimental setups where an agent moves around in an environment and receives local sensations which are aliased in the sense that they do not correspond uniquely to locations in the environment. The environment need not be Euclidean, the agent makes no Euclidean assumptions and does not have access to a map of the environment. If the sensations from the environment are vectors (for example, visual patterns) in a continuous space, they are discretized using a vector quantizer. From a sequence of discretized observations and actions, both of which could be egocentric, an agent has to discover the latent topology of its environment to vicariously evaluate different options for navigation. This is a difficult problem due to the aliasing of the observations and a lack of Euclidean assumptions.

This can be formulated as the problem of learning a latent graph from aliased observations at its nodes. An agent performs a sequence of actions $a_1,\ldots,a_N$ (with discrete $a_n\in\{1,\ldots, N_\text{actions}\}$) in an environment $G$, and as a result of each action, it receives an observation, obtaining the stream $x_1,\ldots,x_N$ (with discrete $x_n\in\{1,\ldots, N_\text{obs}\}$). The goal is to recover the topology of the environment $G$ from sequences of actions and observations.
An environment is defined by a directed multigraph $G\equiv \{V, E\}$ with nodes $V \equiv \{v_1,\ldots v_{N_\text{nodes}}\}$ and edges $E \equiv \{e_1,\ldots e_{N_\text{edges}}\}$. Every node is labeled with a discrete observation. At each time step $n$, the agent will exist at a node and observe $x_n$ to be its label. Multiple nodes can have the same label, so the observation does not identify the node. We use $C(x)$ to refer to nodes with label $x$, also called the \emph{clones} of $x$. When an agent at node $v_i$ executes an action $a$, it will transition to $v_j$ with probability $P(v_j|v_i, a)$. Whenever this probability is larger than 0, an associated directed edge from $v_i$ to $v_j$ is introduced in the graph, labeled with the corresponding action and probability. Note that this means that the graph can contain multiple edges with the same starting and ending node, but labeled with different actions. (This is what makes $G$ a multigraph and not a simple graph). For consistency, all edges originating from the same node and labeled with the same action must have their probabilities sum up to 1.

For a given sequence of actions, $G$ encodes a distribution over sequences, establishing a connection between temporal sequences and arbitrary (not necessarily Euclidean) topologies. To do this effectively, requires a graph learning mechanism that can merge or split contexts appropriately \citep{niv2019learning}. CSCG achieves this by creating a latent space of clones that has the flexibility to split or merge contexts, and having a smooth parameterization of the graph learning problem. Having the latent space allows the model to represent long-term temporal dependencies \citep{cormack1987data}, and gives it the flexibility that is not available to models that purely concatenate temporal context in the observation space. 

\begin{figure}[ht!]
    \centering
    \includegraphics[width=\linewidth]{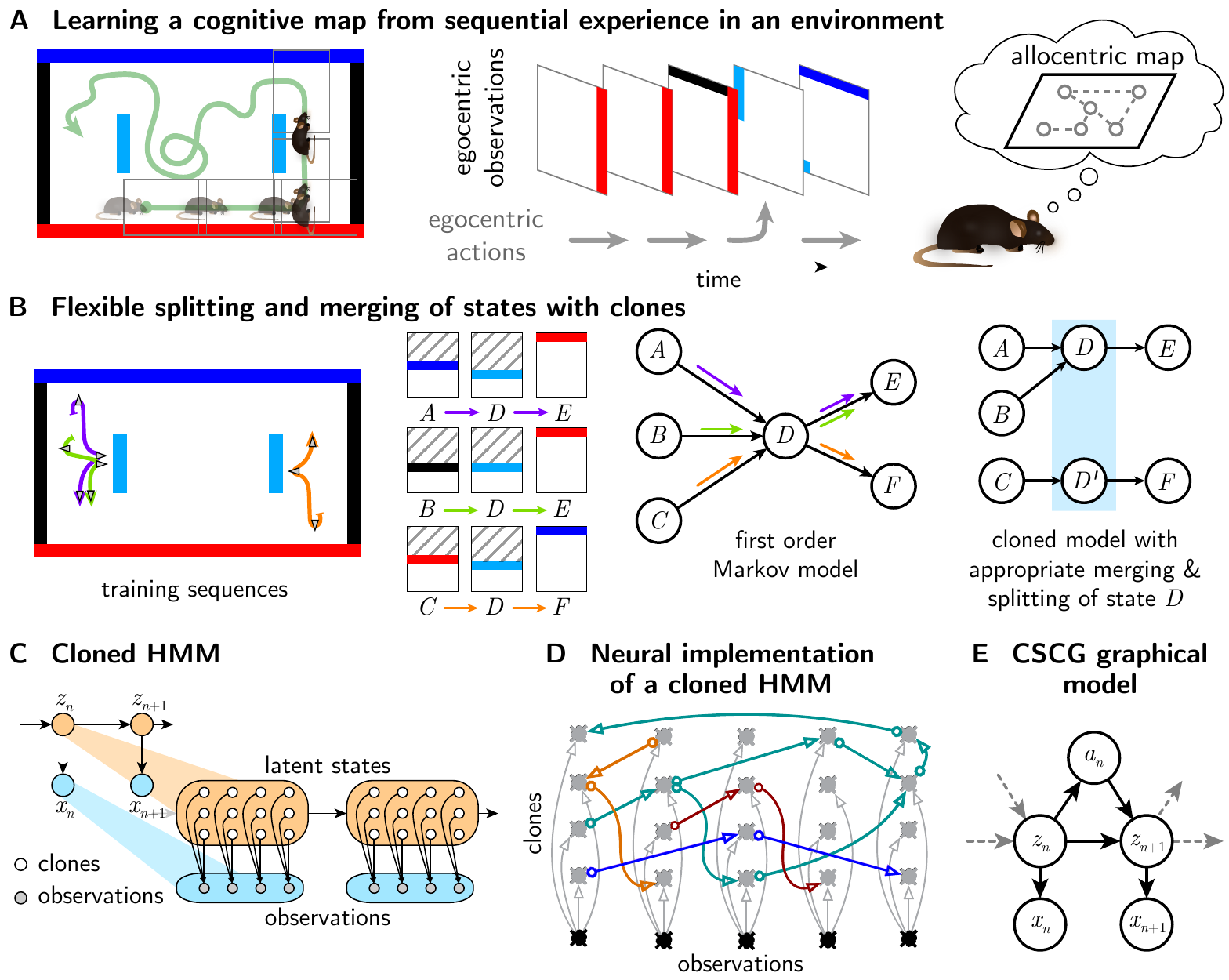}
    \caption{{\bf Clone-structured cognitive graph}. {\bf A}. Learning cognitive maps from sequential sensory observations is challenging because observations  do not identify locations uniquely. {\bf B}. The cognitive map learning problem can be understood as learning a latent graph from observations emitted at every node, where two different nodes can emit the same observation. The challenge is to learn context-specific representations that will disambiguate sensory observations in the latent space. The observation $D$ occurs in three different contexts in sequences $A\rightarrow D\rightarrow E$ (purple), $B\rightarrow D\rightarrow E$ (green), and $C\rightarrow D\rightarrow F$ (orange) from the environment, a distinction that is not represented in a first-order Markov model. Two of these contexts (purple, and green) correspond to the same latent state, and the third (orange) to a different latent state. Cloning $D$ into multiple latent states allows for flexible merging and splitting of contexts as appropriate. {\bf C}. The cloning structure of dynamic Markov coding can be incorporated in an HMM with a structured emission matrix, the cloned HMM. {\bf D}. Neural implementation of a cloned HMM. Neurons in each column are clones of each other that receive bottom-up input from the same observation. Arrow represent axons, and the lateral connections correspond to the cloned HMM transition matrix. Different sequences are in different colors. {\bf E}. Probabilistic model for CSCG which extends cloned HMMs by including actions.}
    \label{fig:CSCG-concepts}
\end{figure}

The above definitions result in a precise, action-conditional probabilistic model for sequences. Using $z_n$ to represent the (unobserved) node at step $n$, and adding a simple per-node policy $P(a_n|z_n)$ to also model the actions, results in the CSCG model. The joint probability of a sequence of observations and actions is
\begin{equation}
    P \lpr x_1, a_1, x_2, \ldots, x_{N-1}, a_{N-1}, x_N\rpr = \sum_{z_1 \in C\lpr x_1\rpr} \ldots \sum_{z_N \in C\lpr x_N\rpr} P\lpr z_1 \rpr \prod_{n=1}^{N-1} P\lpr z_{n+1} | z_n, a_n \rpr
    P\lpr  a_n | z_n \rpr,
\end{equation}
depicted as a graphical model in Fig. \ref{fig:CSCG-concepts}E. Observe that in the action-conditional setting, this corresponds to a hidden Markov model in which the emission matrix is determined by the cloning structure and fixed, which improves its learnability. The model supports causal semantics \citep{pearl2009causality} and learning from interventions \citep{peters2017elements, eaton2007exact}. A learned transition matrix is a directed multi-graph, and reusing this transition matrix and the cloning structure to remap to a new environment can be considered as learning using soft interventions \citep{pmlr-v2-eaton07a}.

Inference and learning in CSCG can be achieved using biologically plausible mechanisms. The clones in CSCG can be represented by an assembly of neurons. Message-passing inference in CSCG is computationally cheap and biologically plausible using simple integrate-and-fire neurons \citep{rao2004bayesian}. Learning is achieved using Expectation Maximization (EM) which maximizes the likelihood of the model using a local update mechanism analogous to spike timing dependent plasticity \citep{nessler2013bayesian, nessler2009stdp}. 

\section{Results}

\begin{table}[htbp]
\centering
\begin{tabular}{l|p{6cm}|p{5cm}}
\hline
\rowcolor{gray!25}
\textbf{Experiment}  & \textbf{Phenomena} & \textbf{Publications} \\
\hline
Geometry changes	& Place field remaps as determined by geometry & \cite{okeefe1996geometric}\\
\hline
Visual cue rotation	& Place field rotates with cue card  & \cite{muller1987effects} \\
\hline
Barrier addition & Place field disruption near barrier & \cite{muller1987effects}\\
\hline
Landmark vector cells	& Place field remaps w.r.t a landmark  & \cite{deshmukh2013influence} \\
\hline
Linear track	& Place field remaps w.r.t start and end of the track  & \cite{sheehan2021compressed} \\
\hline
Directional place fields	& Place field remapping is sensitive to movement direction & \cite{okeefe1996geometric}\\
\hline
Laps on a track	& Event specific rate remapping and lap cells & \cite{sun2020hippocampal}\\
\hline
Four connected rooms	& Place fields are unaffected by closed doors & \cite{duvelle2021hippocampal}\\
\hline
 Two identical rooms	& Place fields are repeated in two rooms & \cite{fuhs2005influence}\\
\hline
  Hairpin maze	& Direction specific repetition of place fields & \cite{derdikman2009fragmentation}\\
\hline
  Room size expansion	& Place fields expand or stretch based on location w.r.t boundaries & \cite{tanni2022state}\\
\hline
\end{tabular}
\caption{List of experiments, their observed phenomena, and related publications.}
\label{table:experiments}
\end{table}
 
We tested the CSCG model in a variety of experimental settings. The first set of experiments investigated the ability of a CSCG to learn latent topologies from perceptually aliased observation sequences, ability to represent multiple maps in the same model, and the ability to transitively stitch global maps from disjoint overlapping experiences.  Furthermore, we investigated the ability of the model to use previously acquired structural knowledge to guide behavior in novel environments. All these properties are important for the performance of an animal. The second set of experiments investigated CSCG's ability to reproduce and explain a broad set of well known experimental phenomena from the hippocampus (see Table \ref{table:experiments}). These phenomena can be broadly divided into spatial, geometry-related, and landmark-related remapping \citep{muller1987effects, okeefe1996geometric, deshmukh2013influence, sheehan2021compressed}, phenomena with both spatial and temporal components \citep{sun2020hippocampal, okeefe1996geometric}, and place field repetition, distortion, and changes with respect to environmental connectivity \citep{fuhs2005influence, derdikman2009fragmentation, duvelle2021hippocampal}. In addition, we performed a set of experiments that serve as testable predictions for CSCG's ability to explain the mechanisms underlying hippocampal phenomena.

\subsection*{CSCG can construct maps from aliased egocentric observations in diverse environments.}

CSCGs are successful in learning latent topologies in a variety of environments, including 2D and 3D layouts (Fig. \ref{fig:CSCG-examples}A and B) and mazes, from purely sequential aliased random walk observations. In the uniform room example in Fig. \ref{fig:CSCG-examples}A, the agent received egocentric visual observations quantized though a vector quantizer, and took egocentric actions, with four possible heading directions in each location. The visible input to the agent depended on its location as well as head direction. Learning in CSCG discovered the latent headings and locations and represented them using separate clones (Fig. \ref{fig:CSCG-examples}A)(iii). Each node in the graph (Fig. \ref{fig:CSCG-examples}A(iii)) corresponds to a clone, and its color represents the local observation it is attached to. Note that the learning of the transition graph discovered $4$ clones per spatial location, this corresponds to the $4$ possible headings that an agent can be in (see ``Methods'' for more details). CSCGs are also able to correctly learn the topology of 3D surfaces (bucky ball, cube) from sequential aliased egocentric observations (Fig. \ref{fig:CSCG-examples}B), correctly inferring the latent local and global loop closures.

While each clone in the transition graph in Fig. \ref{fig:CSCG-examples}A(iii) is `bottom-up' responding to the local sensation indicated by the color, that sensation needs to occur in the latent sequential context specified by the transition graph. By representing sequential contexts in the latent space, these clones come to represent variables like locations and heading that are not directly sensed. An experimenter can obtain the place field of a clone by creating a map representing the arena that the agent is moving in, and marking and accumulating the instantaneous activities of the clone at the present ground-truth location of the agent on that map. Examples of such place fields are shown in Fig. \ref{fig:CSCG-examples}A(iv). The clones in Fig. \ref{fig:CSCG-examples}A(iii) are also head direction sensitive, which corresponds well with the observation in \cite{acharya2016causal} that place fields show head direction sensitivity when they are mapped conditioned on head direction. The head direction sensitivity will be strongest in those locations where the animal has very different visual inputs based on the head direction, compared to the locations in the middle where the animal receives the same visual input in all head directions, consistent with contemporary observations about view sensitivity of place fields \citep{acharya2016causal, Jercog2019-lp, Moore2021-uk, Muller1994-yn}. 
While the place field can give rise to the interpretation that the clone is responding to that particular location, this is purely an interpretive convenience for the experimenter. The agent itself has no mechanism by which it can derive a place field from the activity of its neurons. As we show in the next section, the agent does not need to compute place fields to locate itself, nor need to decode locations from the clones to make navigation decisions. 

\begin{figure}[ht!]
    \centering
    \includegraphics[width=0.95\linewidth]{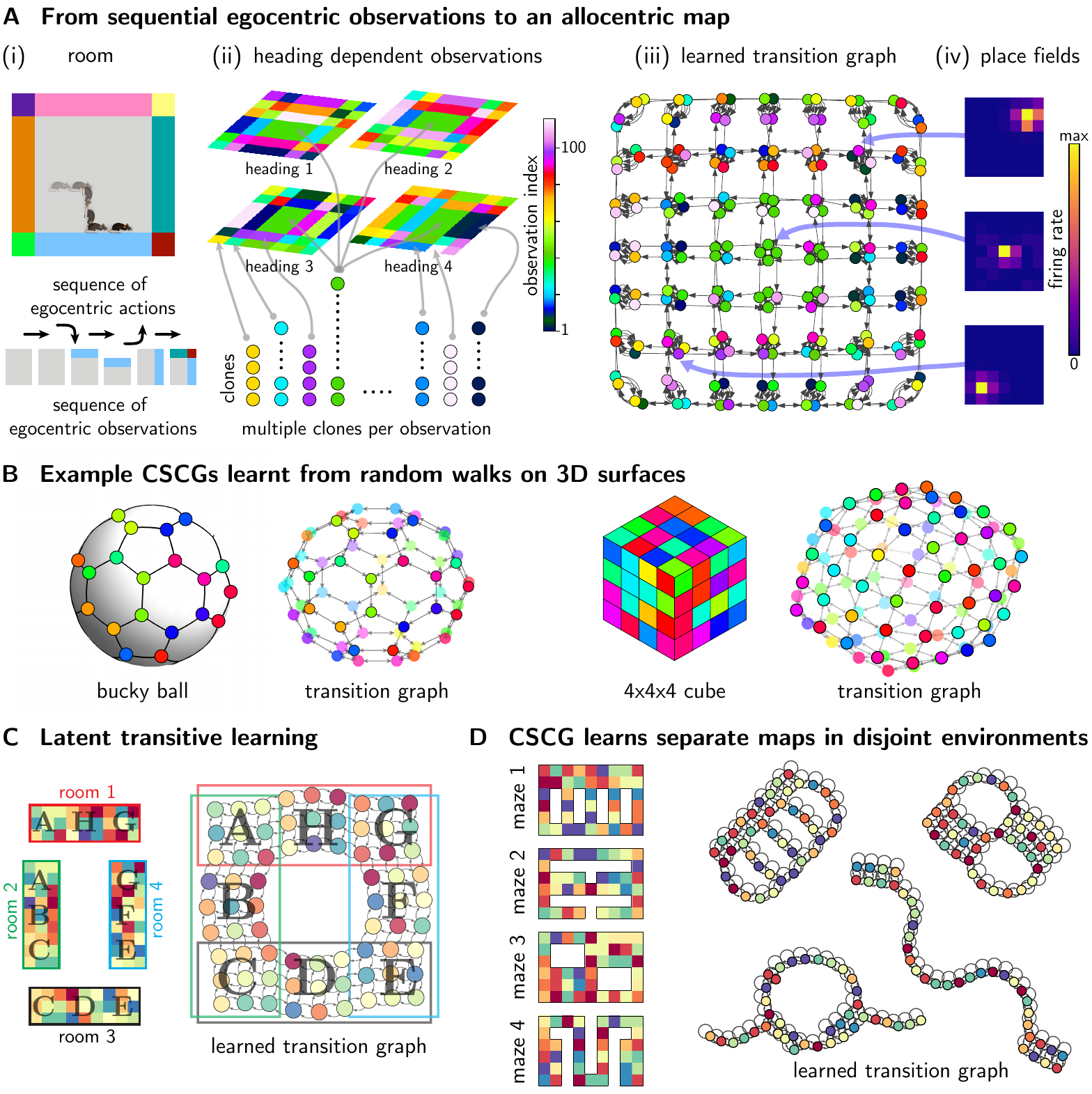}
    \caption{{\bf CSCGs learn diverse latent topologies, transitively stitch them, and transfer structure to new environments}. {\bf A}. CSCGs learn allocentric maps from aliased egocentric local observations from a room with uniform interiors (i) even with long runs of the same observation.  (ii) Each sensation, shown by the color, is attached to a set of latent states (clones) through the emission matrix. Through learning of the transition matrix, these clones learn to represent different temporal contexts of that sensation. (iii) Learned transition graph among clones. Each clone's color represents the observation it is attached to. (iv)  Activations of the clones as the agent navigates the room can be used to compute their place fields, which reveal the spatial locations they represent. {\bf B}. CSCGs are able to correctly learn the topology of 3D surfaces. Shown are the learned transition graphs with node colors indicating the observation that node is connected to. (\emph{continued on next page}) }
    \label{fig:CSCG-examples}
\end{figure}

\begin{figure}[t!]
  \captionsetup{labelformat=adja-page}
  \ContinuedFloat
  \caption{{\bf Properties of CSCGs}. {\bf C}. An agent experiences four overlapping rooms in disjoint sequential episodes. $A, C, E, G$ are the overlapping segments and $B, D, F, H$ are exclusive to the rooms. CSCG learning stitches together the disjoint experience into a coherent global map. {\bf D}. An agent experiences multiple sequential episodes sampled from four non-overlapping mazes. In this case, CSCG learning correctly learns separate maps for each maze.}
\end{figure}

CSCGs make complex latent transitive inferences during learning, and represent the learned information to enable novel transitive inferences \citep{eichenbaum1999hippocampus}. When different overlapping sections of an environment are exposed to the agent in disjoint episodes, CSCGs learn the underlying map that stitches together the whole environment (Fig. \ref{fig:CSCG-examples}C), including the global loop closures. Even though the agent never experienced a transition from locations in segment $A$ to locations in segment $F$ during training, the learned map enables correct inferences between all pairs of such locations,   such as the path from segment $F$ to segment $A$ being shorter via segment $H$ as opposed to via segment $D$.  
When environments are really disjoint, CSCGs learn to separate the maps, and simultaneously represent multiple maps in memory without being explicitly instructed about map boundaries during training (Fig. \ref{fig:CSCG-examples}D). The appropriate map can then be recalled as hidden state inference \citep{sanders2020hippocampal}, and used to guide behavior.

\begin{figure}[ht!]
    \centering
    \includegraphics[width=\linewidth]{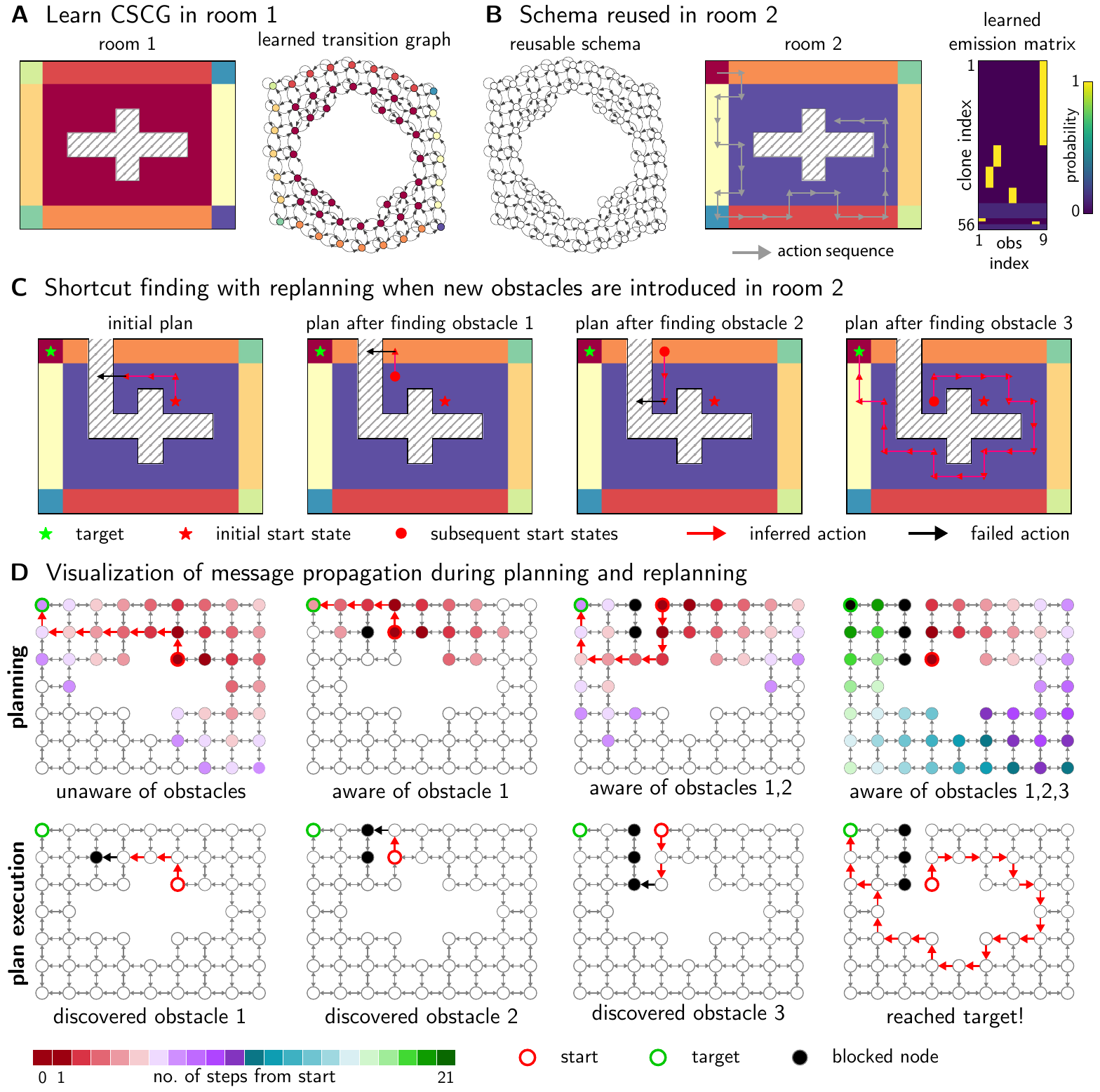}
    \caption{ {\bf Shortcut finding in a novel room}. The learned transition graph of a CSCG trained on one room ({\bf A}) can be considered a reusable schema. Given partial observations in a second, novel room with an identical layout ({\bf B}), the CSCG can re-use the previously learned latent structure to rapidly navigate around obstacles and find the shortest path to a target ({\bf C}). ({\bf D}) The graphs in the top row are a visualization of message propagation during planning and re-planning. Messages propagate outward from the starting clone. The first plan is unaware of the obstacles, and the agent discovers an obstacle only when the action sequence is executed (bottom row) and a planned action fails (at a node in black). This initiates re-planning from the new location, and the new plan routes around the obstacle. }
    \label{fig:Shortcut-finding}
\end{figure}

\subsection{Replay-based planning and schema-based transfer enable shortcut inference in dynamic settings}
A behaving agent can keep track of its state as the most likely clone given past observations, without having to invoke any concepts about space or place fields. If the agent wants to navigate to a remembered goal of a visual sensation, the action sequences that achieve this can be inferred directly by clamping the corresponding clones and propagating messages. If the environment is learned correctly, this inference process is exact, and will recover the sequence of actions that will take the agent from the current state to the desired goal. Message-passing based planning in CSCGs is akin to replays in the hippocampus \citep{olafsdottir2018role}. A striking advantage of CSCG in comparison to models that purely predict, is that learned maps can be quickly reconfigured to reflect changes in the environment. When a previously passable route is blocked, the corresponding structural modification can be made in the latent graph, and message-passing based inference will utilize this updated information about the environment to navigate around obstacles.

CSCGs can also transfer prior knowledge to new environments and infer novel shortcut paths through unobserved locations by treating the learned transition graph as a schema \citep{baraduc2019schema, barry2006boundary} and learning just the emission matrix. 
To demonstrate this ability, we first trained a CSCG using aliased observations from a random walk in a room (room 1, Fig. \ref{fig:Shortcut-finding}A). Next, we placed the agent in an unfamiliar room with the same structure (room 2, Fig. \ref{fig:Shortcut-finding}B). As the agent walks in the new room, we keep the transition matrix of the CSCG fixed and update the emission matrix with the EM algorithm. Just by walking along the periphery of room 2, the CSCG is able to infer the shortest path between visited locations through previously unvisited locations. Further, if we block the path with obstacles and a planned action fails, the CSCG is able to initiate replanning at the blocked location and reroute to the target (Fig. \ref{fig:Shortcut-finding}C). Thus, even with partial knowledge of a novel room, an agent can vicariously evaluate the sequence of actions to be taken to reach a destination by reusing the CSCG’s transition graph from a similar, previously experienced room.

\subsection*{Remapping due to changes in overall geometry, visual cues, or landmarks can be explained using sequence learning}

Changing the interpretation of place fields from explicitly representing spatial locations to representing the sequential context in which a sensation occurs explains a wide variety of place cell remapping phenomena. In the transition graph in Fig. \ref{fig:CSCG-examples}A(iii), each state should be interpreted, not as responding to a specific location in the room, but as responding to the specific sequence of observations leading up to that location. As we demonstrate, sequential interpretation of spatial representations can explain a variety remapping driven by changes in geometry, visual cues, transparent or opaque barriers, landmarks, distances to start or end locations etc. 

\begin{figure}[ht!]
    \centering
    \includegraphics[width=\linewidth]{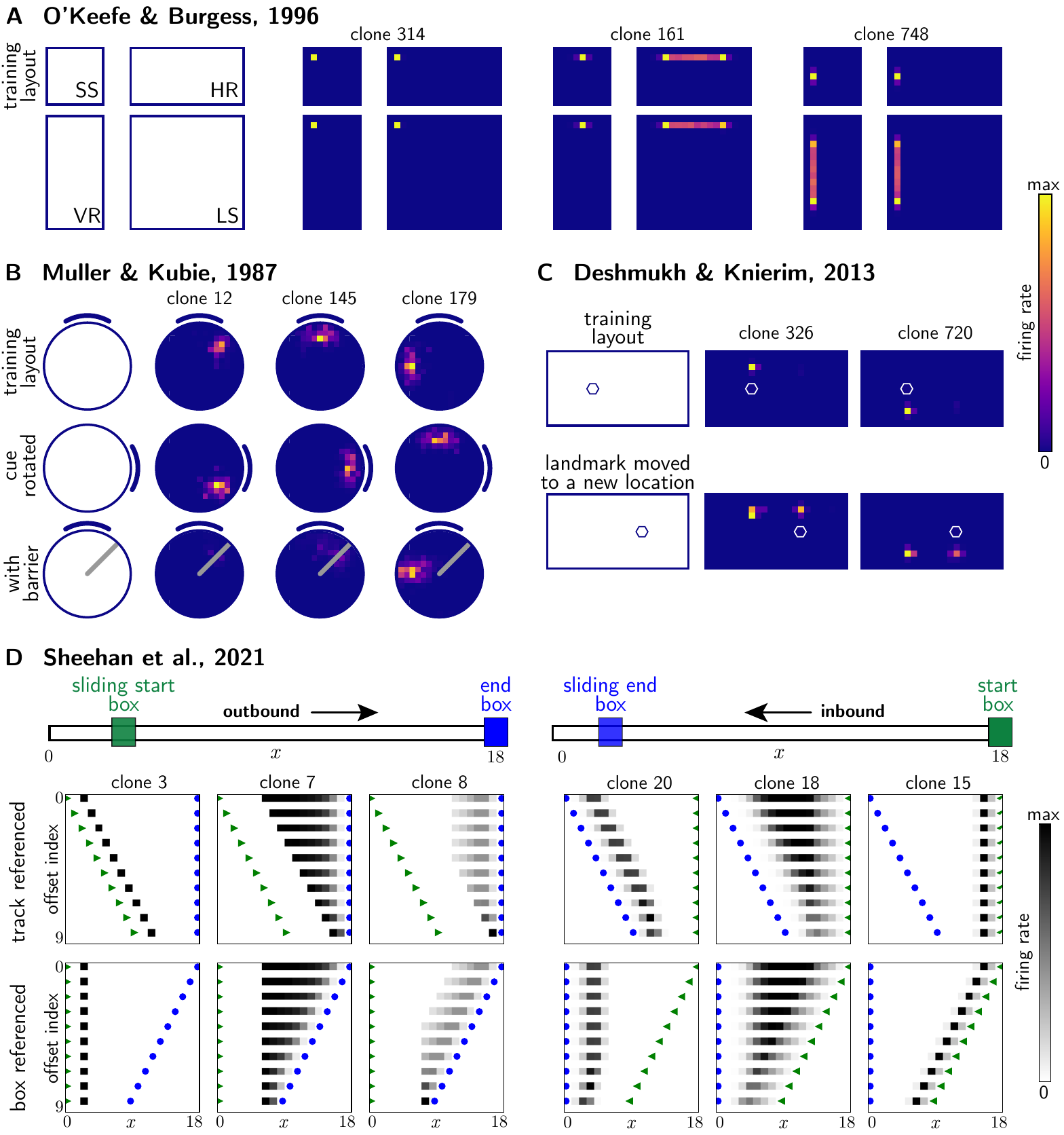}
    \caption{{\bf CSCG can reproduce several place field remapping phenomena}. {\bf A}. Geometric determinants of place fields, \cite{okeefe1996geometric}. A CSCG was first trained using a random walk in the small square (SS) room. Place fields were computed in the training room as well as the horizontal rectangle (HR), vertical rectangle (VR), and large square (LS) rooms. For clone 314, the place field is anchored to the top-left corner and stays the same in all four rooms because the place field of this neuron is determined by the unique visual input at the boundary. Place fields anchored to an edge of the room, however, split into two components when the room is elongated along the corresponding axis - clones 161 and 748 along the horizontal and vertical axes, respectively. (\emph{continued on next page}) }
    \label{fig:PF-remapping}
\end{figure}

\begin{figure}[t!]
  \captionsetup{labelformat=adja-page}
  \ContinuedFloat
  \caption{{\bf CSCG can reproduce several place field remapping phenomena}. {\bf B}. Effects of changes in the environment on the spatial firing of hippocampal place cells, \textcite{muller1987effects}. A CSCG was first trained using a random walk in a circular room with a cue at the 12 o' clock position. Place fields were then computed in the training room, the room with the cue rotated by $\ang{90}$, and the room with a barrier introduced. For each clone, we observe that the place fields also rotate by $\ang{90}$ when the cue is rotated. For most clones, the place fields completely disappear when the barrier is introduced. An exception is clone 179, where the place field in the original layout is sufficiently far away from the barrier. {\bf C}. Landmark vector cells, \textcite{deshmukh2013influence}. A CSCG was trained in a rectangular room with a landmark (depicted by a hexagon) on one side of the room. Place fields were then computed in the training room as well as a modified room in which the landmark was moved to a different location. For the modified layout, we observe that the place fields now have two components - one at the same location as the place field in the original layout, and another at the same vector displacement from the new location of the landmark. {\bf D}. A compressed representation of spatial distance in the rodent hippocampus, \textcite{sheehan2021compressed}. A CSCG was trained on a linear track of length 18 steps using outbound (left to right) and inbound (right to left) walks. Place fields were computed using trials with different starting positions for the outbound trajectories and different end positions for the inbound trajectories. The top and bottom rows correspond to place fields in the reference frame of the track and the start box, respectively. We observe that most place cells coded for distance from the starting box. A few clones, e.g., clones 8 and 20, are anchored to the end box. Further, clones have gradually widening fields with distance from the starting or ending locations. 
  }
\end{figure}

The classic experiments on geometry-change-driven remapping \citep{okeefe1996geometric} can be explained using CSCGs as follows: changing the geometry of a room changes the locations where similar sequential contexts will be observed. In these experiments, place fields that developed while the rat trained in an arena remapped in a geometry-dependent manner when the arena was elongated or widened. We demonstrate this by first training a CSCG on a small square (SS) room (size $9\times9$)
and uniform interior and observing the place field changes of clones in test rooms that varied in size along the two dimensions (see Fig \ref{fig:PF-remapping}A). The activations of clones in a CSCG represents the posterior distribution over latent states given the past sequence of observations. As described earlier, the specific sequential context in which clones activate can be interpreted as coding for location. Since the interiors of a uniform room have undifferentiated local sequential context, the responses of clones in the center will be anchored with respect to the boundaries because of the relative uniqueness of the observations there. When navigating an elongated room using the CSCG learned from the smaller room, the internal states will reliably signal end-of-room states when the agent is near the boundaries of the new room. This effectively creates two loci for sequential contexts. The same clone that fired in the sequential context corresponding to a specific location in the original room will now fire at two different locations due to the splitting of the sequential contexts in the elongated room, as reflected in the remapped responses of clones 161 and 748 in Fig. \ref{fig:PF-remapping}A. In contrast, the response of  clone 314 does not remap and remains the same in all four rooms. This is because this neuron's sensory input already includes part of the boundary, and also because the sequence it represents has shorter undifferentiated segments from the boundary, making it strongly anchored. Although these results were originally characterized as boundary-vector coding, our results show that the major findings of \cite{okeefe1996geometric} can be explained using sequence representation without using geometric concepts. As we describe later, the sequence perspective also naturally explains the temporal dependence of the remapped place fields. Of course, with further training in the new environment, the remapping will diminish because new place fields representing the new environment will develop with more experience in that environment. 

The classic Muller and Kubie experiments \citep{muller1987effects} showing a variety of remapping phenomena can also be explained using CSCG, which we illustrate in Fig. \ref{fig:PF-remapping}B. In these experiments, rats were trained in a circular arena with a cue card placed on the wall. Researchers found a variety of remapping phenomena with respect to rotation of the cue card and introduction of opaque or transparent barriers. To investigate these phenomena, we first trained a CSCG in a circular arena with a cue card at the 12 o' clock position. In this environment, the differentiated sequential contexts will develop with reference to the cue card. When we computed place fields with this CSCG in an arena where the cue was rotated, the place fields also rotated accordingly because they are always referenced to the context and not the absolute location. Placing a barrier in the arena has two effects that destroy the place field for some clones. One effect is that the barrier prevents the agent from taking some trajectories that are important for revealing the relevant sequential contexts for some clones. The second is that the presence of the barrier can change the visual sensation in its vicinity. Both these effects combine to explain why place fields are disrupted when a barrier is placed through its center, and not affected when the barrier is far away.

CSCGs also explain why place cells can be seen as encoding a vector relationship to local landmarks \citep{deshmukh2013influence}. Just like cue cards, or boundaries, landmark objects placed in an environment act as disambiguating contexts with respect to which sensations at other locations are encoded. Thus, when a landmark is moved, some of sequential contexts also move in reference to that landmark. We illustrate this landmark vector remapping phenomenon in Fig. \ref{fig:PF-remapping}C. We first trained a CSCG in a rectangular layout with a landmark on one side of the room. We computed place fields in this layout as well as a modified version in which the landmark was moved to a different location. In the modified layout, the place fields now have two components - one at the same location as in the original layout, and the second at the same relative displacement from the new location of the landmark.   

In more recent experiments \citep{sheehan2021compressed}, rats were trained on outbound and inbound traversals on a linear track that could be changed in length. Responses to the appropriate sequential contexts in a CSCG naturally explain the remapping of place fields observed as the track length varies. To demonstrate this, we first trained a CSCG on a linear track of length 18 steps using both outbound (left to right) and inbound (right to left) walks. We then computed place fields separately on outbound and inbound trajectories, for various track lengths (Fig. \ref{fig:PF-remapping}D). We observed that most clones coded for distance from the starting position. The place fields gradually widened with distance from the starting position reflecting the growing uncertainty in the distance from the starting point. There were also clones anchored to the end point of the trajectories. 

\subsection*{Sequence representation can explain puzzling phenomena that mix spatial and temporal effects}

Sequential contexts naturally explain the direction sensitivity of place field remapping reported in \textcite{okeefe1996geometric}.
When the room is elongated, some place fields that were unimodal in the original room remapped to produce two peaks, corresponding to two subcomponents in the elongated room. It was observed that these peaks were direction sensitive: the left subcomponent was active during rightward travel and vice versa. We tested CSCG for the same effects using the same settings as in Fig. \ref{fig:PF-remapping}A, by plotting the fields conditioned on the direction of travel. In the HR room, rightward and leftward trajectories of the agent strongly activated the left and right peaks of the place field, respectively, as shown in Fig. \ref{fig:directional-rfs-lap-experiment}A. This is because only one of the sequential contexts that activate a clone occurs in a directional walk, which is a natural consequence of representing locations using sequential contexts. In contrast, a purely geometric model like the boundary vector model \citep{barry2006boundary} does not offer an explanation for the direction sensitivity of place field remapping.

\begin{figure}[ht!]
    \centering
    \includegraphics[width=\linewidth]{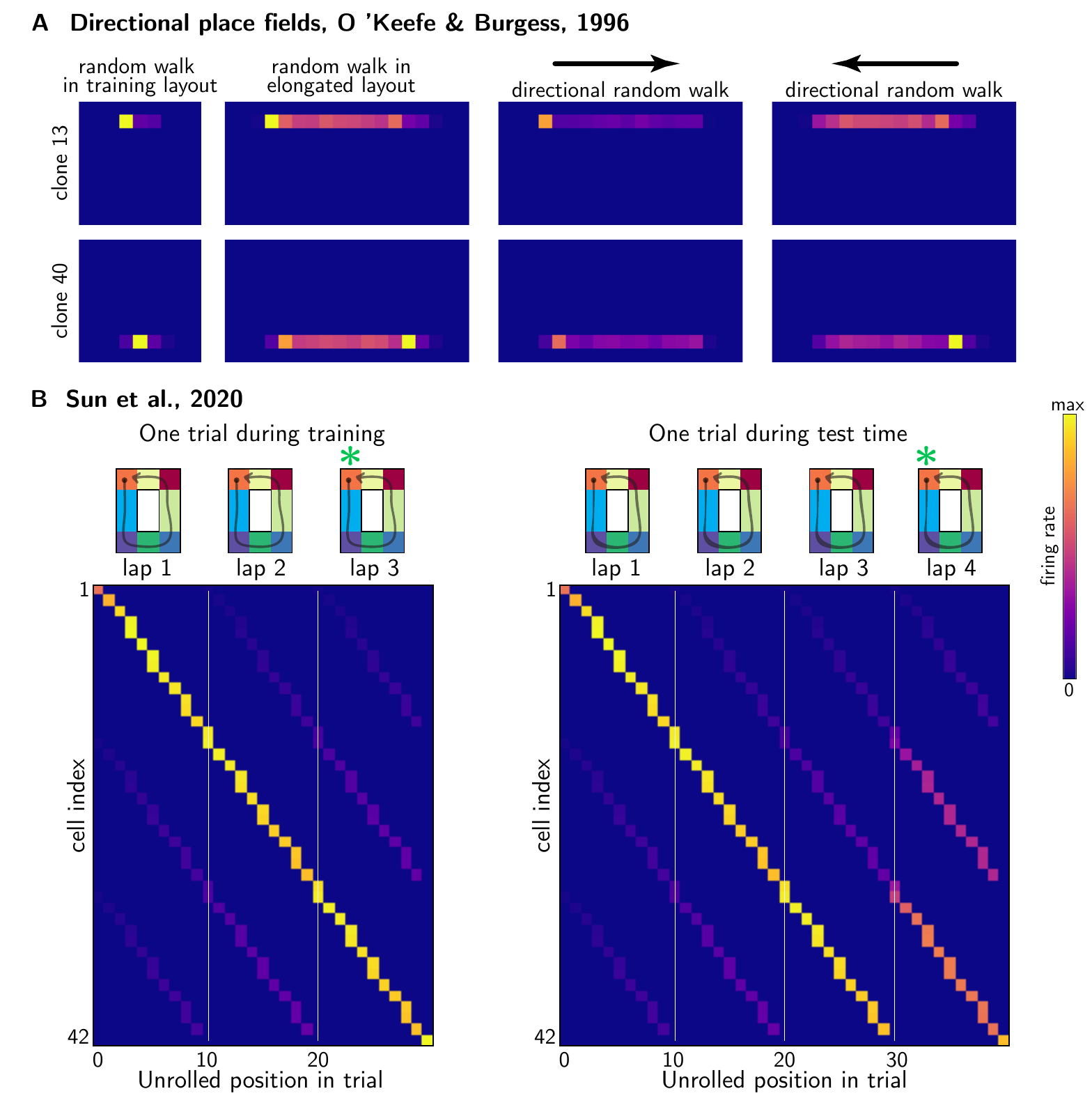}
    \caption{{\bf A}. We reproduce the directional place fields reported by \textcite{okeefe1996geometric}. We first trained a CSCG in a square room. In an elongated rectangular room, place fields of clones of the CSCG elongate with two strong peaks (second column from the left). The left peak is located such that its distance from the left wall is the same as in the square room. The same holds true for the right peak relative to the right wall. Further, when we use directional random walks to compute place fields, the two independent subcomponents corresponding to these peaks are revealed. A particular subcomponent is stronger when the walk starts from the wall to which the location of the subcomponent is tied: left and right components in the $\rightarrow$ and $\leftarrow$ walks, respectively. 
    (\emph{continued on next page})
    }
    \label{fig:directional-rfs-lap-experiment}
\end{figure}

\begin{figure}[t!]
  \captionsetup{labelformat=adja-page}
  \ContinuedFloat
  \caption{{\bf B}. Lap-neurons and event-specific representations. A CSCG was trained on observations from  laps in a rectangular maze similar to \textcite{sun2020hippocampal}. The training sequence consisted of three laps followed by a reward state (green $\ast$) at the end. We also considered test trials with four laps where the reward state was at the end of the fourth lap. We show the place fields computed using the training and test trials, respectively. Rows correspond to clones. The place fields on the training trials show that there are different clones that are maximally active for different laps. But most clones are also partially active at their corresponding location in other laps, similar to the neurophysiological observations in \cite{sun2020hippocampal}. For the test trials, we observe that the lap three clones are significantly active in both the third and the fourth laps. This additional activation shifted precisely by one lap reflects the fact that the third lap is no longer rewarded and that an extra lap is needed to receive a reward.
  }
\end{figure}

CSCG can also explain recently discovered phenomena like event-specific rate remapping (ESR) cells \citep{sun2020hippocampal}, which signal a combination of location and lap number for different laps around a maze, without postulating special coding mechanisms. Fig. \ref{fig:directional-rfs-lap-experiment}B shows a similar setting to an experiment in \textcite{sun2020hippocampal} where a rat runs multiple laps in a looping rectangular track before receiving a reward. We trained a CSCG on trials comprising three laps of a rectangular track with a reward state at the end of the third lap. A CSCG exposed to the sequence of observations from such trials learned to distinguish the laps and to predict the reward at the end of the third lap, without the help of any explicit lap-boundary markers in the training sequence. This is reflected in the place fields of the clones for the training trials (left panel in Fig. \ref{fig:directional-rfs-lap-experiment}B) - each clone is maximally active for an observation when it occurs in its specific lap. However, each clone also shows weak activations when its corresponding observation is encountered in other laps, a signature of ESR. This occurs naturally in the CSCG due to smoothing and the inference dynamics. CSCGs can also explain the remapping of ESR cells. We computed place fields on test trials comprising four laps, instead of three, in which the the reward was at the end of the fourth lap. The lap three clones were strongly activated in both the third and the fourth laps, reflecting the change in when the reward state is reached (right panel in Fig. \ref{fig:directional-rfs-lap-experiment}B). 
 
\subsection{CSCG can predict what kinds of changes in the environment lead to remapping}

CSCGs show that environmental connectivity changes need not lead to place field remapping even when the agents' behavior shows adaptation to the change, a phenomenon that researchers found puzzling.  
In \textcite{duvelle2021hippocampal}, rats ran in a 4-room maze where the doors connecting the rooms could be selectively locked to change the connectivity of the arena. The agent's behavior reflected that it recognized the connectivity changes of the environment, but the place fields did not remap in response to these connectivity changes. The authors found this lack of remapping puzzling and argued that place cells do not encode a topological map. However, CSCGs show that place cells can encode global location in their activations, global topology in the cell-to-cell connectivity, and still not show remapping in response to the manipulations in \textcite{duvelle2021hippocampal}.

To demonstrate this, we trained a CSCG using a random walk in an environment comprising four square rooms that are connected by two-way doors, similar to the experimental setting in \textcite{duvelle2021hippocampal}. Each room had visual cues that distinguished it from the other rooms. CSCG learned the global topology of the arena in the transition matrix, and the activation of clones corresponded to locations, as in previous experiments (Fig. \ref{fig:PF-repetition}A top row). We then tested for two environmental modifications used in \textcite{duvelle2021hippocampal} - (i) one door was locked both ways effectively creating a blockade, and (ii) all doors were locked in one way allowing only an anti-clockwise direction of traversal in the environment. The corresponding modifications were made in the CSCG transition matrix by modifying the connections appropriately, and planning routes in this modified CSCG corresponded to the reported successful navigation. We then computed place fields using the appropriately modified CSCGs paired with the arena connectivity changes, and compared these to the fields from the original CSCG in the original arena. In Fig. \ref{fig:PF-repetition}A, we show that the place fields were the same across all three settings, consistent with the observations in \textcite{duvelle2021hippocampal}. 

The reason for lack of remapping can be understood by realizing that the connectivity change blocked paths without any change in the visual cues. The blocked path affected only a few of the potential sequences that were responsible for that place field, a change that is too small to be reflected in the aggregated sequential responses. However, the connectivity change can still lead to large changes in behavior, for example, in navigation between the two rooms. In CSCGs, those changes will be reflected in the replay messages used for planning, and in the computed shortest paths.

\begin{figure}[ht!]
    \centering
    \includegraphics[width=\linewidth]{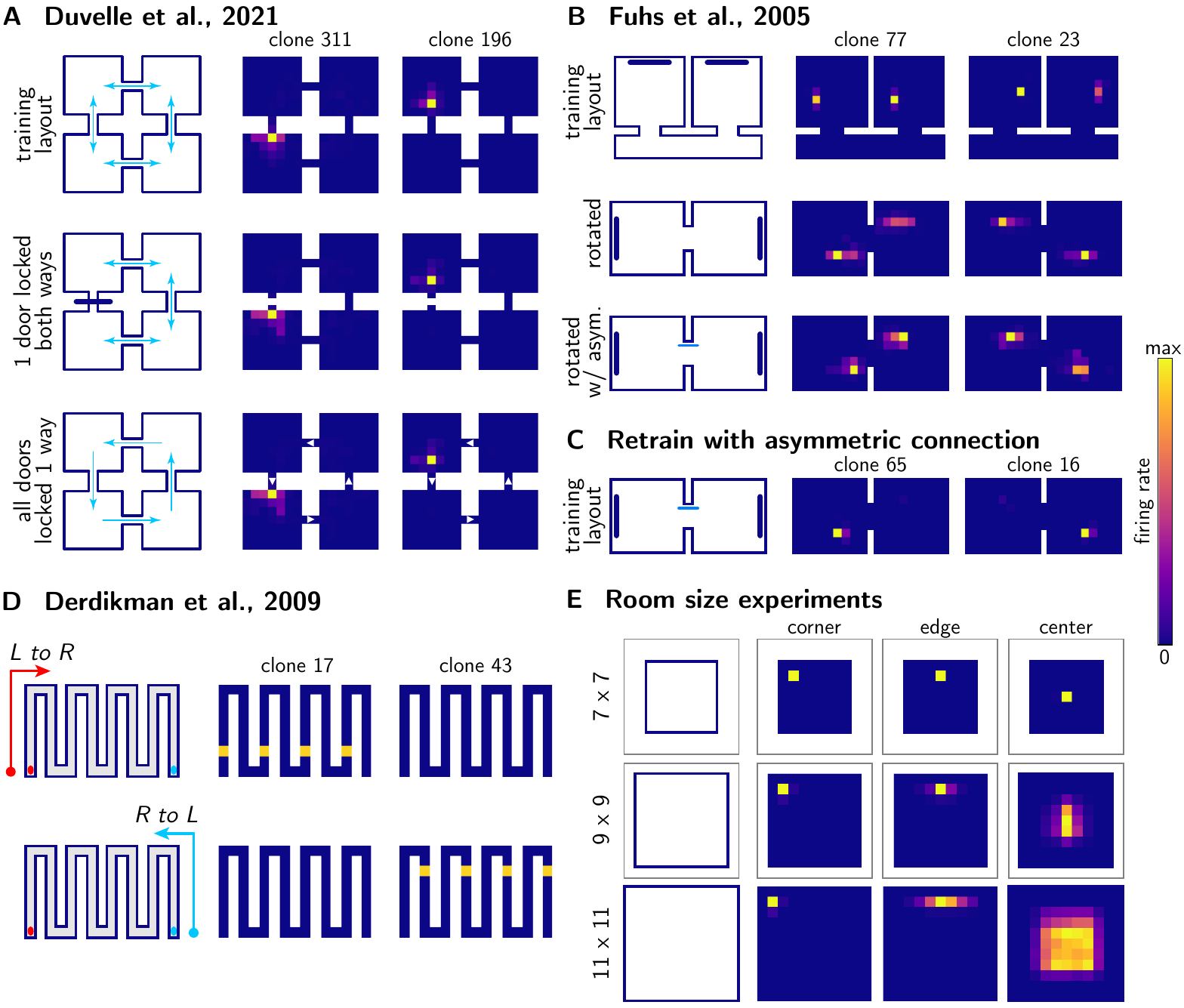}
    \caption{ {\bf CSCG can reproduce various observations about place cells such as place field repetition and size, shape variations.} {\bf A}. Hippocampal place cells encode global location but not connectivity in a complex space, \textcite{duvelle2021hippocampal}. A CSCG was first trained in a layout comprising four square rooms that are connected by two-way doors. Place fields were computed in the training layout and two modified layouts - (i) one door locked both ways, and (ii) all doors locked one way. We observe that place fields remain the same across all three settings. {\bf B}. Place field repetition in visually identical environments. A CSCG was first trained on a layout comprising two identical rooms in the same orientation connected by a corridor. The behavior of the CSCG was then studied in a layout where the two rooms were abutted by rotating them such that their orientations differed by $\ang{180}$. We also considered a second modification, where we introduced an asymmetery in the connection between the abutting rooms. In all these three layouts, we observed place field repetition across the two rooms. Note that these results are incongruent with those in \textcite{fuhs2005influence}, where they observed place field repetition only in the same orientation layout. {\bf C}. However, when we retrain a CSCG in the opposite orientation layout with asymmetric connectivity, we observe that the place field repetition disappears. (\emph{continued on next page})
    }
    \label{fig:PF-repetition}
\end{figure}

\begin{figure}[t!]
  \captionsetup{labelformat=adja-page}
  \ContinuedFloat
  \caption{{\bf CSCG can reproduce various observations about place cells such as place field repetition and size, shape variations}. {\bf D}. Direction dependent place field repetition in a hairpin maze, \textcite{derdikman2009fragmentation}. A CSCG was trained using $L\rightarrow R$ and $R\rightarrow L$ walks on a hairpin maze. Place fields were then computed using only $L\rightarrow R$ (top row) or $R\rightarrow L$ (bottom row) walks. We observe direction dependent repetition of place fields. {\bf E}. Place field size and shape as a function of room size, \textcite{tanni2022state}. We trained three different CSCGs on square rooms, with uniform interiors, of side length $7, 9$ and $11$, respectively. The capability of a CSCG to learn the map of a room with uniform interior degrades as the room gets larger. This is reflected in the place fields. Place fields that are anchored to the corner of the room retain their size and shape across sizes. However, place fields at the edge and center of the room elongate and become larger as the room size increases. 
  }
\end{figure}

\subsection*{Sequence learning explains place field repetition, size and shape variations.}

Place fields distort along the boundaries, and increase in size systematically towards the center of an empty arena \citep{tanni2022state}. In very elongated rooms, place fields have multiple lobes. In some settings, place fields are known to repeat in identical rooms \citep{skaggs1998spatial, fuhs2005influence}. While all these phenomena appear to be spatial, CSCGs provide cogent explanations for these in terms of sequence learning: all of them result from state aliasing due to the difficulty in creating different latent states for temporal contexts that are identical for long number of steps.

 To demonstrate place field repetition in visually identical environments, we trained a CSCG in a layout comprising two visually identical rooms in the same orientation and connected by a corridor, as shown in Fig. \ref{fig:PF-repetition}B, similar to the setting in \textcite{fuhs2005influence}. Place fields computed in this layout show repetition, i.e., clones are active at the same location in both rooms. We also considered a layout in which the two rooms were abutted by rotating them such that their orientations differed by $\ang{180}$. In \textcite{fuhs2005influence}, it was reported that place field repetition disappears in the different-orientation setting. This was attributed to the rats potentially being able to maintain their inertial angular orientation. With CSCGs, in the absence of an external ``compass'', we observe that place field repetition persists in the modified layout, even after the introduction of an asymmetric connection between the two rooms. However, when the CSCG was retrained on the different-orientation setting with an asymmetric connection between the rooms, it was able to partially split contexts in the two rooms. This resulted in unique place fields for most clones, as show in Fig. \ref{fig:PF-repetition}C. If the sensory input to CSCG is augmented with an external head direction input, then the different orientation setting results in unique place fields in CSCGs, similar to what is observed in \textcite{fuhs2005influence}. 

In Fig. \ref{fig:PF-repetition}D, we reproduce the direction-dependent place field repetition reported in \textcite{derdikman2009fragmentation}. We trained a CSCG on a hairpin maze, with distinct end markers, using left to right $\lpr L \rightarrow R\rpr$ and right to left $\lpr R \rightarrow L\rpr$ walks.  
Place fields computed using this CSCG using only $L \rightarrow R$ or $R \rightarrow L $ walks reveal direction dependent place field repetition, as shown in Fig. \ref{fig:PF-repetition}D. For example, clone 17 is activated at the same location in all segments of the maze, but only in the $L \rightarrow R$ traversal. The two ends of the maze have different observations, which provides the CSCG enough context to disambiguate the two directions of travel. However, for each direction of traversal, the observations are the same in all segments of the maze resulting in the repetition of place fields. 

To study the effect of room size on place fields \citep{tanni2022state}, we trained three different CSCGs on square rooms, with uniform interiors, of side length 7, 9 and 11, respectively. As an agent moves away from the boundaries to the center of an empty room, different sequential trajectories start to look the same, making it difficult for the learning algorithm to split the contexts into different clones. This results in the same clone representing more contexts than it would in the periphery of the room where contexts can be easily distinguished. In place field mapping, this will appear as an enlargement of the place fields in the center of the room (``center'' column in Fig. \ref{fig:PF-repetition}E). Similarly, the observations along the edge of a room might not all develop into distinct clones, resulting in multiple observations along the edge being aliased into the same clone. This aliasing, due to the elongation of the same evidence, will appear as an elongation of the place field (``edge'' column in Fig. \ref{fig:PF-repetition}E). 

Place field size expansion \citep{tanni2022state} in an empty arena happens because of the same reason as place field repetition in two identical iso-oriented rooms. Both can be explained by the inability of the model to split very long-term temporal contexts into distinct latent clones with the given amount of training. (Of course longer training will partially overcome this problem, which is observed in animals as well.) In that sense, larger place fields are the same as place field repetition, just happening in adjacent locations.

\begin{figure}[ht!]
    \centering
    \includegraphics[width=\linewidth]{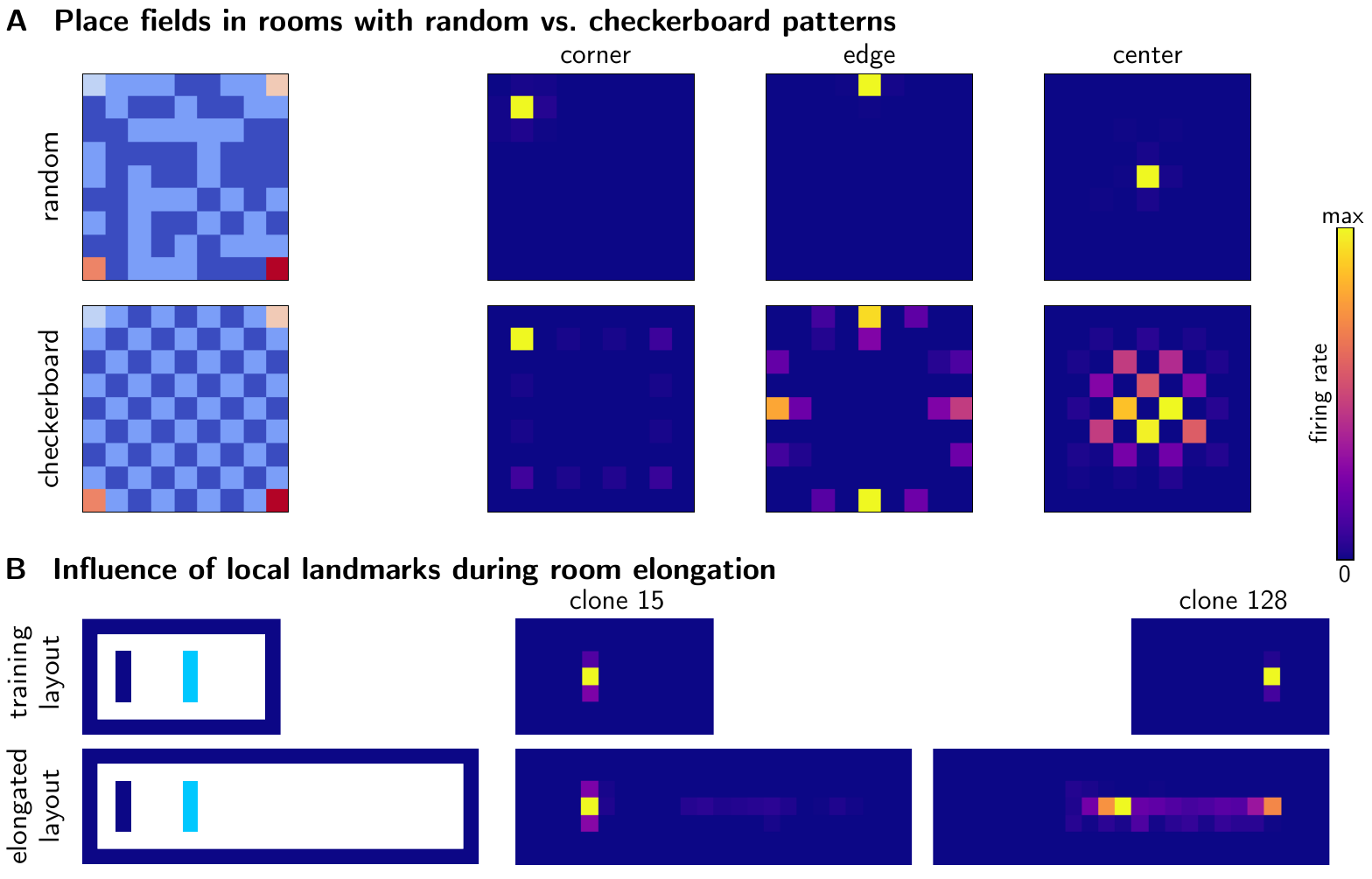}
    \caption{{\bf CSCG based predictions about place fields}. {\bf A}. What controls place field change is not the rate of visual field change, but the uniqueness of the visual context. We observe that a CSCG trained on the checkerboard room has more expanded place fields compared one trained on a room with a random pattern. {\bf B}. Influence of local landmarks during a room elongation experiment. We first trained a CSCG on a rectangular room with local landmarks on one side of the room. We computed place fields in the training room and an elongated room in which the landmarks are in the same position as the training room. Place fields anchored to the landmark (clone 15) stay the same in both layouts, but place fields sufficiently far from the landmarks expand in the elongated layout (clone 128).
    }
    \label{fig:PF-predictions}
\end{figure}

\subsection*{Novel experiments and predictions}
CSCGs can also make experimentally testable predictions for yet to be observed phenomena. One such prediction is the following. What controls how place fields change is not the rate of visual change, but the uniqueness of the visual context. To demonstrate this, we trained two CSCGs on square rooms with checkerboard and random patterns on the floor, respectively. We observed that the place fields in the checkerboard room were more expanded, as shown in Fig. \ref{fig:PF-predictions}A. This is because the same context repeats throughout the interior of the room, making it difficult for the learning to split the contexts into different clones. 

CSCGs provide a mechanistic explanation for the question of when and why do place fields globally or partially remap? The answer: place cell responses are driven by their sequential contexts, and changes that significantly affect the sequential context of a neuron is what determines when and how its field will remap.   Any change that makes the same sequential context occur in different parts of the room, will result in that field partially appearing in the new place. The organization and specificity of local context driving the responses of a cell will have a significant impact on its remapping. A cell that is tuned to sequences in the middle of a uniform room, will have its place fields anchored by the boundaries that are relatively more unique, causing the field to remap when when the boundaries are moved. However, if the cell had some other local cues, for example markings on the floor, that would provide it a unique sequential context, then the cell's field will not remap when the boundary is moved. In Fig. \ref{fig:PF-predictions}B, clone 15 and clone 128 are two cells from the CSCG trained in the training layout. When the the room is elongated, the place field of clone 128 remaps as shown. This is because the local sequential context for clone 128 was anchored by the cyan marking on the left and the boundary on the right. These partial contexts occur in two different place in the elongated room. In contrast, the local sequential context for clone 15 is anchored by the blue marking on its left and the cyan marking on its right, and those did not change when the room was elongated. This means, locally, clone 15 will see the same sequential contexts after room elongation, resulting in a lack of remapping in its place field. 

\section{Discussion}

The discovery of place cells is a striking success of hippocampus research, and place field mapping has served as a valuable tool in revealing the representational properties of neurons in the HPC. However, anomalies have been accumulating over the simple view that place cells represent just locations \citep{acharya2016causal, sun2020hippocampal, buzsaki2017space}. Place fields distort around boundaries, and split along trajectories. They are direction sensitive \citep{acharya2016causal, Moore2021-uk}, and can even represent the lap count in running loops \citep{sun2020hippocampal}. In some cases, insertion of a boundary in between a place field clearly disrupts the place field, suggesting that fields are related to the connectivity of the underlying environment \citep{muller1987effects}. Yet, place fields can remain unchanged when the connectivity of the environment changes without any visible cues \citep{duvelle2021hippocampal}. If the environmental change is not reflected in place fields, how are the rats able to change their behavior in the new environment? In summary, many of these questions about the role of place cells -- what they represent, how those representations are learned, how they are used, and how they change with respect to environmental manipulations remain unanswered in the location-centric description of hippocampal neurons.

In this paper, we pursued the strong hypothesis that the hippocampus performs a singular algorithm that learns a sequential, relational, content-agnostic structure of its environment \citep{buzsaki2017space, Dabaghian2014-ix}, and demonstrated evidence for its validity. Our learning model, the CSCG, inverts the observation stream to learn a latent generative model that is producing the stream of sensations. With CSCGs, we demonstrated how a vast variety of experimental observations about hippocampal place cells can be explained by the singular key insight that spatial representation is an emergent property of latent higher-order sequence learning. 
We demonstrated this by first showing that pure temporal learning is sufficient to acquire cognitive maps that have locations, space, heading etc. We also showed how such latent graphs can transfer knowledge across environments. And finally, we showed that multiple phenomena that we observe are natural byproducts of sequence learning and inference, without having to directly model the phenomena itself.

While CSCG draws up on many past and contemporary models of hippocampus \citep{uria2022model}, it is significantly different in many aspects. In contrast to temporal context models (TCM) \citep{howard2002distributed} that accumulate sequential context in the observation space, the sequential representation in CSCG is in the latent space, giving it the ability to model more complex and long duration temporal dependencies. The ability of CSCG to represent locations as sequences crucially depends up on having a latent representation. Although successor representations \citep{stachenfeld2017hippocampus} can model temporal relations, they are not directly applicable in the aliased settings we consider here, and do not learn spatial representations from egocentric sensory inputs.  Contemporary work on Tolman-Eichenbaum machines (TEM) \citep{whittington2018generalisation} have many similarities to CSCG in inspiration. However, unlike CSCG, TEMs do not learn latent graphs in aliased settings like ours. Instead, TEM focuses on learning general transitivity rules applicable to a single graph from multiple noisy realizations of that graph. Moreover, TEMs do not deal with multiple graphs at the same time \citep{Sanders2020-sy}, or do latent transitive stitching. In the context of learning spatial representations, TEMs have so far been demonstrated only in allocentric settings. More importantly, TEM is formulated purely as a predictive model and its internal representation does not learn a modifiable graph that corresponds to the environment. Therefore, TEM doesn't have the same ability as CSCGs to deal with dynamic environments quickly by changing its graph connectivity, or to form hierarchies through community detection \citep{schapiro2016statistical} on the latent graph. 

Unlike other computational models of place fields, CSCGs do not use grid fields to learn place fields and still explain varied remapping phenomena. Recent experimental evidence suggests that grid cells are not necessary for learning \citep{Tan2017-er,Brandon2014-tv} and continued functioning of place cells \citep{Brandon2014-tv}. If grid cells outputs are available, CSCG can utilize those as additional sensations. This would speed up learning in the middle portions of empty arenas where unique sensations are not available, and it will also help stabilize the place fields away from the boundaries or other landmark cues \citep{Mallory2018-vk, Muessig2015-pg}, consistent with the idea of grid cells providing an optional scaffolding for place cells \citep{mulders2021structured}. 

The most important message from our work is that many of the diverse fascinating hippocampal phenomena might be artifacts of Euclidean place field mapping. Hippocampal cells are usually interpreted by plotting their responses on to a 2D map corresponding to the environment, collapsing the sequential responses in to a static place field. Characterizing place field maps in terms of Euclidean concepts is akin to characterizing the effects rather than the underlying causes, and might be the source of new phenomena. 
Often these new phenomena are explained away invoking familiar, but ultimately unsatisfactory, answers like distributed coding, or mixed selectivity. These answers are unsatisfactory because instead of answering the questions they just shift the questions elsewhere. Our experiments show that phenomena that look extremely different -- for example place-field expansion in a uniform room and event-specific responses in a lap running -- can have the same underlying explanation which can be understood through the sequence learning model. We hope this opens up a new avenue of exploration that takes us away from the familiar questions centered on encoding and decoding locations. 

Much remains to be explored on this new path we have struck out on. We have only briefly touched up on replay based planning, and schemas, and both can be expanded in future research. Our work can also be expanded in the direction of active learning and inference. Reward mechanisms can be layered on top of CSCG. CSCGs have the ability for temporal abstractions via community detection \citep{schapiro2016statistical} on the underlying graphs, an idea worthy of more exploration. Current models are learned using random walks. Potentially efficient exploration techniques can be developed as active learning on CSCGs. Most importantly, we hope our work gives a concrete tool that would help hippocampal researchers think beyond the place field paradigm.

\section*{Methods}

\subsection*{Expectation-Maximization learning of CSCGs}

Cloned Hidden Markov Models (HMMs), first introduced in \cite{dedieu2019learning}, are a sparse restriction of overcomplete HMMs \citep{sharan2017learning} that can overcome many of the training shortcomings of dynamic Markov coding \citep{cormack1987data}. Similar to HMMs, cloned HMMs assume the observed data $x_1, \ldots, x_N$ are generated from a hidden process $z$ that obeys the Markovian property
\begin{equation*}
    P(x_1, \ldots, x_N, z_1, \ldots, z_N) = 
    P(z_1) \prod_{n=1}^{N-1} P(z_{n+1} | z_n) \prod_{n=1}^{N}P(x_n|z_n)
\end{equation*}
Here $P(z_1)$ is the initial hidden state distribution, $P(z_{n+1} | z_n)$ is the transition probability from $z_n$ to $z_{n+1}$, and $P(x_n | z_n)$ is the probability of emitting $x_n$ from the hidden state $z_n$.

In contrast to HMMs, cloned HMMs assume that each hidden state maps deterministically to a single observation. Further, cloned HMMs allow multiple hidden states to emit the same observation. All the hidden states that emit the same observation are called the {\it clones} of that observation.

CSCGs build on top of cloned HMMs by augmenting the model with the actions of an agent. In this section we first review the expection-maximization learning of cloned HMMs, before describing the learning of CSCGs.

\subsubsection*{Expectation-Maximization learning of Cloned HMMs}

The standard algorithm to train HMMs is the expectation-maximization (EM) algorithm \citep{wu1983convergence}, which in this context is known as the Baum-Welch algorithm.
Learning a cloned HMM using the Baum-Welch algorithm requires a few simple modifications: the sparsity of the emission matrix can be exploited to only use small blocks of the transition matrix both in the Expectation (E) and Maximization (M) steps.

Learning a cloned HMM requires optimizing the vector of prior probabilities $\pi$: $\pi_u = P(z_1 = u)$ and the transition matrix $\mathbf{T}$: $\mathbf{T}_{uv} = P(z_{n+1} = v | z_n=u)$. To this end, we assume the hidden states are indexed such that all the clones of the first emission appear first, all the clones of the second emission appear next, etc. Let $N_{\textrm{obs}}$ be the total number of emitted symbols. The transition matrix $\mathbf{T}$ can then be broken down into smaller submatrices $\mathbf{T}(i, j), i,j \in \{1, \ldots, N_{\textrm{obs}} \}$. The submatrix $\mathbf{T}(i, j)$ contains the transition probabilities $P(z_{n+1} | z_n)$ for $z_n\in C(i)$ and $z_{n+1} \in C(j)$, where $C(i)$ and $C(j)$ correspond to the hidden states (clones) of emissions $i$ and $j$ respectively. 

The standard Baum-Welch equations can then be expressed in a simpler form in the case of cloned HMM. The E-step recursively computes the forward and backward probabilities and then updates the posterior probabilities. The M-step updates the transition matrix via row normalization.

\textbf{E-Step}
\begin{align*}
\alpha(1) &= \pi(x_1)  &
\alpha(n+1)^\top &= \alpha(n)^\top \mathbf{T}(x_n, x_{n+1})\\
\beta(N) &= 1(x_N)  & \beta(n) &= \mathbf{T}(x_n, x_{n+1}) \beta(n+1)
\end{align*}
\begin{align*}
\xi_{ij}(n) &=  \frac{\alpha(n)\circ \mathbf{T}(i, j) \circ \beta(n+1)^\top}{\alpha(n)^\top \mathbf{T}(i, j) \beta(n+1)}\\
\gamma(n) &=  \frac{\alpha(n) \circ \beta(n)}{\alpha(n)^\top\beta(n)}.
\end{align*}

\textbf{M-Step}
\begin{align*} 
\pi(x_1) &= \gamma(1)\\
\mathbf{T}(i, j) &=  \Big(\sum_{n=1}^N \xi_{ij}(n)\Big) \oslash \Big(\sum_{j=1}^{N_{\textrm{obs}}} \sum_{n=1}^N \xi_{ij}(n)\Big).
\end{align*}
where $\circ$ and $\oslash$ denote the element-wise product and division, respectively (with broadcasting where needed). All vectors are $M\times1$ column vectors, where $M$ is the number of clones per emission. We use a constant number of clones per emission for simplicity here, but the number of clones can be selected independently per emission. Cloned HMMs exploit the sparsity pattern in the emission matrix when performing training updates and inference, and achieve significant computational savings when compared with HMMs.

\subsubsection*{CSCGs: Action-augmented cloned HMMs}

CSCGs are an extension of cloned HMMs in which an action happens at every timestep (conditional on the current hidden state) and the hidden state of the next timestep depends not only on the current hidden state, but also on the current action. The joint probability density function on the observations and the actions is given by
\begin{equation*}
P(x_1,\ldots,x_N, a_1,\ldots,a_{N-1}) =  \sum_{z_1 \in C(x_1)} \ldots \sum_{z_n \in C(x_n)}  P(z_1) 
\prod_{n=1}^{N-1} P(z_{n+1}|z_n, a_n) P(a_n | z_n)
\end{equation*}
and the standard cloned HMM can be recovered by integrating out the actions. 

We group the actions with the next hidden state to remove loops, and create a chain that is amenable to exact inference. In other words, we rewrite the joint probability density function as
\begin{equation*}
    P(x_1,\ldots,x_N, a_1,\ldots,a_{N-1}) =  \sum_{z_1 \in C(x_1)} \ldots \sum_{z_n \in C(x_n)}  P(z_1) 
\prod_{n=1}^{N-1} P(z_{n+1}, a_n|z_n)
\end{equation*}

Learning a CSCG requires optimizing the vector of prior probabilities $\pi$: $\pi_u = P(z_1 = u)$ and the action-augmented transition matrix $\mathbf{T}$: $\mathbf{T}_{uvw} = P(z_{n+1} = v, a_n = w | z_n=u)$. Similar to cloned HMMs, we can break the action-augmented transition matrix $\mathbf{T}$ into smaller submatrices $\mathbf{T}(i, k, j), i,j \in \{1, \dots, N_{\textrm{obs}}\}, k \in \{1, \dots, N_{\textrm{actions}}\}$. The submatrix $\mathbf{T}(i, k, j)$ contains the transition probabilities $P(z_{n+1}, a_n = k | z_n)$ for $z_n\in C(i), z_{n+1} \in C(j)$, where $C(i)$ and $C(j)$ correspond to the hidden states (clones) of emissions $i$ and $j$ respectively. All the previous considerations about cloned HMMs apply to CSCGs and the EM equations for learning are also very similar:

\textbf{E-Step:}
\begin{align*}
\alpha(1) &= \pi(x_1)  &
\alpha(n+1)^\top &= \alpha(n)^\top \mathbf{T}(x_n, a_n, x_{n+1})\\
\beta(N) &= 1(x_N)  & \beta(n) &= \mathbf{T}(x_n, a_n, x_{n+1}) \beta(n+1)
\end{align*}
\begin{align*}
\xi_{ikj}(n) &=  \frac{\alpha(n)\circ \mathbf{T}(i, a_n, j) \circ \beta(n+1)^\top}{\alpha(n)^\top \mathbf{T}(i, a_n, j) \beta(n+1)}\\
\gamma(n) &=  \frac{\alpha(n) \circ \beta(n)}{\alpha(n)^\top\beta(n)}.
\end{align*}

\textbf{M-Step:}
\begin{align*} 
\pi(x_1) &= \gamma(1)\\
\mathbf{T}(i, k, j) &=  \sum_{n=1}^N \xi_{ikj}(n) \oslash \sum_{k=1}^{N_{\textrm{actions}}}  \sum_{j=1}^{N_{\textrm{obs}}} \sum_{n=1}^N \xi_{ikj}(n).
\end{align*}

In \cite{George2021-qt}, it was observed that the convergence of EM for learning the parameters of a CSCG can be improved by using a smoothing parameter called the pseudocount. The pseudocount is a small constant that is added to the accumulated counts statistics matrix ($\sum_{n=1}^N \xi_{ikj}(n)$), which ensures that any transition under any action has a non-zero probability. This ensures that the model does not have zero probability for any sequence of observations at test time. 

\subsection*{Egocentric actions and observations}

To demonstrate how CSCGs can reproduce various experimental findings regarding the hippocampus, we consider experimental setups where an agent performs egocentric actions and makes egocentric observations. We assume that the agent is exploring a layout on an axis-aligned grid. At each time step, the agent can perform one of three actions: (i) go forward by one step along its current heading, (ii) turn left by $\ang{90}$, and (iii) turn right by $\ang{90}$. Note that the agent's heading at any given time can only be one of four headings, which we denote by the four cardinal directions.

\begin{figure}[ht!]
    \centering
    \includegraphics[width=\linewidth]{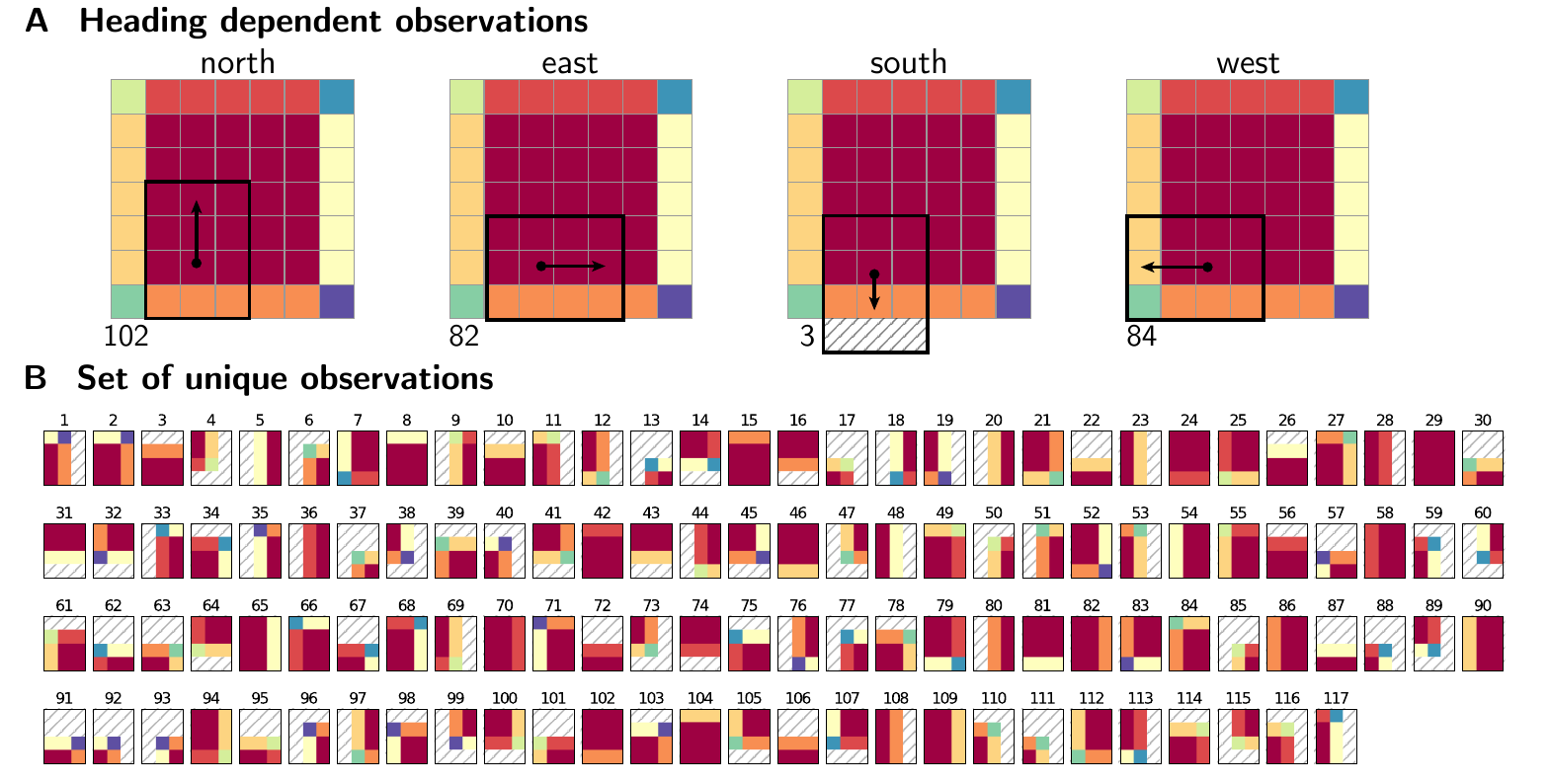}
    \caption{{\bf Egocentric observations for an example $7 \times 7$ square layout.} {\bf A}. The four heading dependent observations at the location denoted by the black dot, labeled by the corresponding observation indices.  The observation window is of size $f_l = 4$, $f_w = 3$. {\bf B}. The set of all unique observations for this example layout and their corresponding indices. The slashed areas correspond to regions outside the layout that are not visible to the agent.}
    \label{fig:Egocentric_observations}
\end{figure}

\begin{figure}[ht!]
\centering
\includegraphics[width=\linewidth]{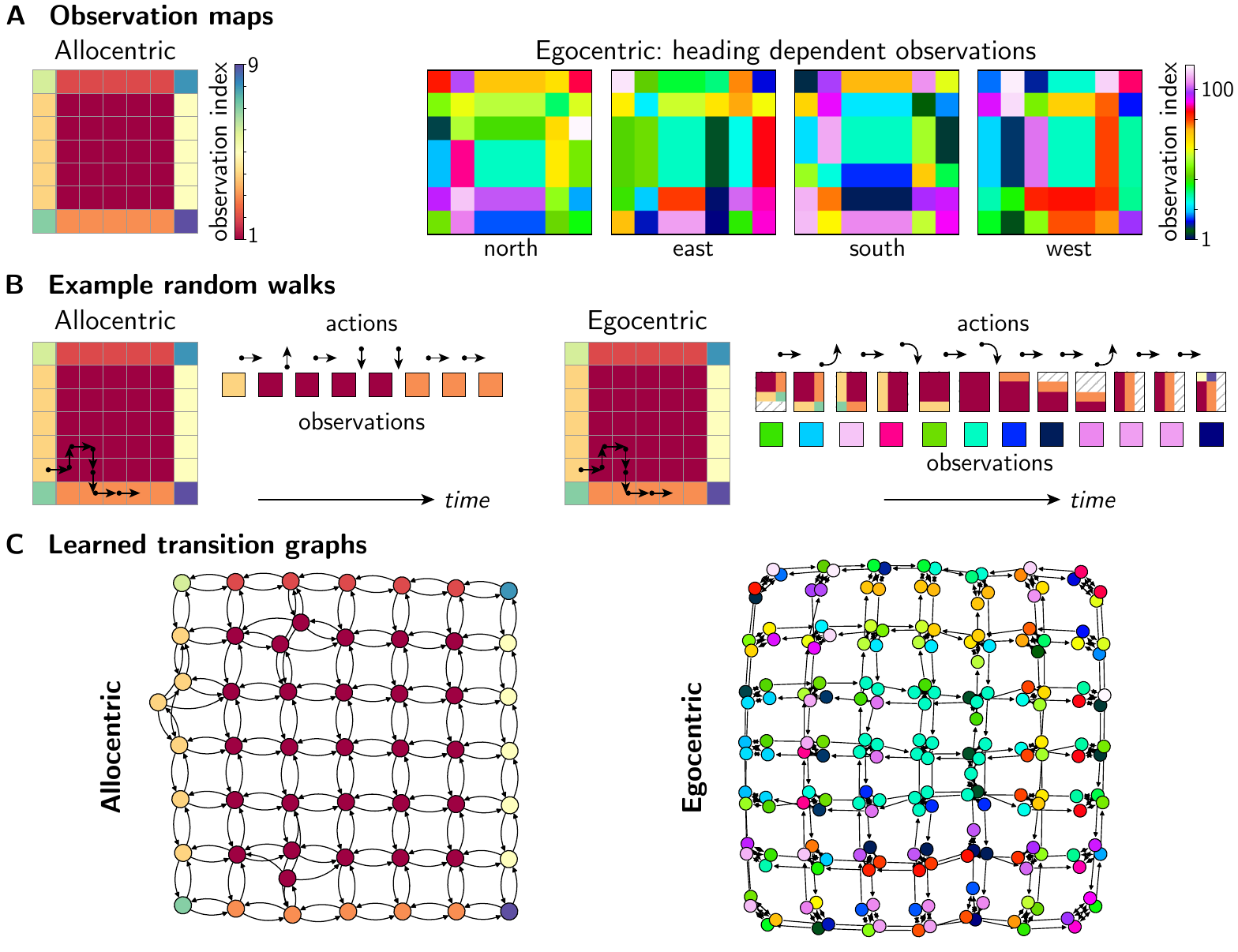}
\caption{{\bf Allocentric vs. egocentric settings.} {\bf A}. Observation maps for an example $7 \times 7$ square layout in the allocentric (left) and egocentric (right) settings. In the egocentric case,
the maps specify the observation index at any given position and heading, for a field of view of size $f_l = 4$, $f_w = 3$. {\bf B}. The sequence of observations and actions for an example trajectory in the two cases. For the egocentric case, the observations include the actual $4 \times 3$ observation patches (top row) and their corresponding indices (bottom row). {\bf C}. The learned transition graphs for CSCGs trained on the example layout using allocentric (left) and egocentric (right) settings. In both cases, the learned graphs correctly correspond to a $7 \times 7$ grid.}
\label{fig:Allocentric-vs-Egocentric}
\end{figure}

The observation of the agent at any given time depends on its position in the layout and its current heading. We assume that the agent has a field of view of length $f_l$ and width $f_w$. This field of view is such that the agent can see up to $f_l-1$ steps in front and $1$ step behind it, and is symmetric along the width axis. Fig. \ref{fig:Egocentric_observations}A shows the four heading dependent observations at the location marked by the black dot in an example $7 \times 7$ square layout. Inaccessible or invisible regions in the field of view are marked by gray slashes. Fig. \ref{fig:Egocentric_observations}B shows the set of all possible observations of size $(f_l=4, f_w=3)$ for this example layout. Each possible observation is also assigned a label/index. Fig. \ref{fig:Allocentric-vs-Egocentric}A (right panel) shows the observation index at each position, for all four possible headings, in the example layout.     

\subsubsection*{CSCGs in allocentric vs. egocentric settings}
For some of the results used to highlight the properties of a CSCG, we used allocentric actions and observations. In this setting, at each time step, the agent can perform one of four actions: go (i) left, (ii) right, (iii) up, or (iv) down. The observation at each time step is just the color/index of the current location in the layout.

Here, we compare CSCGs trained using observations and actions in allocentric vs. egocentric settings. We use a square layout of size $7 \times 7$, as shown in Fig. \ref{fig:Egocentric_observations}A. For the egocentric setting we use a field of view of size $f_l = 4, f_w=3$. The observation maps for both settings are shown in Fig. \ref{fig:Allocentric-vs-Egocentric}A. In the allocentric case, the no. of unique observations $N_{\textrm{obs}} = 9$. In contrast, $N_{\textrm{obs}} = 117$ for the egocentric setting. Fig. \ref{fig:Allocentric-vs-Egocentric}B illustrates the observation and action sequence for an example trajectory in both settings. 

In each setting, we collected a stream of $50,000$ action-observations pairs. We allocated $50$ clones per observation, set the pseudocount to $5 \times 10^{-4}$, and ran EM for a maximum of $1000$ iterations to train a CSCG.  We then used Viterbi decoding to identify the relevant states/clones that are in use. Fig. \ref{fig:Allocentric-vs-Egocentric}C shows the learned transition graphs of the CSCGs learned in the two settings. In both cases, each node in the graph corresponds to a clone and its color corresponds to the observation the clone emits. The edges correspond to non-zero probability transitions between clones. In the allocentric case, each spatial location is represented by one clone in the graph. On the other hand, in the egocentric case, each location is represented by four clones corresponding to four possible headings. Importantly, a CSCG is able to learn a graph that correctly represents the topology of the environment in either setting, without any spatial information as inputs during learning.   

\subsection*{Place fields of clones in CSCGs}
Given a sequence of observations and actions, we define the activation of a clone $i$ at time $n$ as the following marginal posterior probability,
\begin{equation*}
    r_i(n) = P(z_n = i|x_1,\ldots, x_n, a_1, \ldots, a_{n-1})
\end{equation*}
Since the CSCG model (with the action $a_{n-1}$ and hidden state $z_{n}$ collapsed in a single variable) forms a chain, inference on it using belief propagation is exact. The marginal posterior distribution can be computed at each time step as follows,
\begin{equation*}
\bar{r}(n) = \lpr \mathbf{T}\lpr a_{n-1} \rpr^\top r\lpr n-1 \rpr \rpr \circ \lpr E\tilde{x}_n \rpr; \; \; \; r(n) = \frac{\bar{r}(n)}{\sum_{i=1}^{N_{\textrm{clones}}} \bar{r}_i(n)}
\end{equation*}
where $\bar{r}(n)$ is the unnormalized distribution, $\tilde{x}_n$ is a one-hot encoding of the observation $x_n$, $\mathbf{T}\lpr a_{n} \rpr$ is the transition matrix corresponding to action $a_n$, and $E$ is the clone-structured emission matrix. 
The unnormalized marginal distribution at the first time step is computed as $\bar{r}(1) = \pi \circ \lpr E\tilde{x}_1 \rpr$. In settings where the observations are noisy or uncertain, $\tilde{x}_n$ can be a distribution over all possible observations. 

To compute place fields, we first obtain activations of the clones from $N_{\textrm{trials}}$ random walks of the agent, each of length $N_{\textrm{seg}}$, in an environment. We can then use the agent's ground truth spatial information to compute the average activation of a clone at each spatial location in the environment, thus obtaining its place field.  

\section{Acknowledgements}
We thank Matt Botvinick, Kimberly Stachenfeld, Dharshan Kumaran, Charles Blundell, Murray Shanahan and  Demis Hassabis  for critically reading this manuscript and for insightful discussions.

\bibliography{main}

\end{document}